\documentclass[10pt,letterpaper,conference]{IEEEtran}
\IEEEoverridecommandlockouts 

\makeatletter
\newcommand{\AddInputPath}[1]{%
  \ifx\input@path\@undefined
    \def\input@path{#1}
  \else
    \g@addto@macro{\input@path}{#1}
  \fi
}
\makeatother
\AddInputPath{{../}}

\usepackage{etex}

\usepackage[font=footnotesize,caption=false]{subfig} 

\usepackage{relsize}

\usepackage{array}
\usepackage{booktabs,tabularx}
\usepackage{multirow}
\usepackage[binary-units=true,per-mode=symbol,detect-mode=true]{siunitx}
\usepackage{graphicx}

\usepackage[T1]{fontenc}
\usepackage{textcomp}
\usepackage[utf8]{inputenc}
\usepackage[final]{microtype}
\usepackage{icomma}
\usepackage{xspace}

\usepackage[tbtags]{amsmath}
\usepackage{amssymb,amsfonts,bm}
\usepackage{mathtools} 
\usepackage{dsfont}
\usepackage{mathrsfs}
\usepackage{accents}
\usepackage{empheq}
\usepackage{nccmath}
\usepackage{balance}

\usepackage{color}
\usepackage{calc}
\usepackage{tikz}
\usepackage{pgfplots,pgfplotstable}

\usepackage{pdftexcmds}
\makeatletter
\newcommand{\strequal}[2]{\pdf@strcmp{#1}{#2}==0}
\makeatother

\usepackage[capitalize]{cleveref}
\usepackage{refcount}

\usepackage[inline]{enumitem}
\usepackage{algorithm}
\usepackage{algpseudocode}
\makeatletter
\newcommand{\algmargin}{\the\ALG@thistlm}
\makeatother
\newlength{\whilewidth}
\algdef{SE}[parWHILE]{parWhile}{EndparWhile}[1]
  {\parbox[t]{\dimexpr\linewidth-\algmargin}{%
     \hangindent\whilewidth\strut\algorithmicwhile\ #1\ \algorithmicdo\strut}}{\algorithmicend\ \algorithmicwhile}%
\algnewcommand{\parState}[1]{\State%
  \parbox[t]{\dimexpr\linewidth-\algmargin}{\strut #1\strut}}

\makeatletter
\newcommand\fs@spaceruled{\def\@fs@cfont{\bfseries}\let\@fs@capt\floatc@ruled
  \def\@fs@pre{\vspace{.05in}\hrule height.8pt depth0pt \kern2pt}%
  \def\@fs@post{\kern2pt\hrule\relax}%
  \def\@fs@mid{\kern2pt\hrule\kern2pt}%
  \let\@fs@iftopcapt\iftrue}
\makeatother

\usepackage{glossaries}
\usepackage{ifthen}
\usepackage{cite}
\usepackage{multibib}

\usepackage{comment}
\usepackage{todonotes}
\let\legacytodo\todo
\newcommand{\ruggedtodo}[2][]{\tikzexternaldisable\legacytodo[#1]{#2}\tikzexternalenable}
\renewcommand{\todo}[1]{\ruggedtodo[inline]{#1}}

\makeglossaries

\newacronym{leo}{LEO}{low earth orbit}
\newacronym{iot}{IoT}{internet of things}
\newacronym{irs}{IRS}{intelligent reflective surface}
\newacronym{socp}{SOCP}{second-order cone program}
\newacronym{soc}{SOC}{second-order cone}
\newacronym{dsl}{DSL}{digital subscriber line}
\newacronym{wsee}{WSEE}{weighted sum energy efficiency}
\newacronym{mmwave}{mmWave}{millimeter wave}
\newacronym{dfg}{DFG}{Deutsche Forschungsgemeinschaft}
\newacronym{haec}{HAEC}{Highly Adaptive Energy-Efficient Computing}
\newacronym{hpc}{HPC}{High Performance Computing}
\newacronym{mac}{MAC}{multiple-access channel}
\newacronym{bc}{BC}{broadcast channel}
\newacronym{siso}{SISO}{single-input single-output}
\newacronym{simo}{SIMO}{single-input multiple-output}
\newacronym{miso}{MISO}{multiple-input single-output}
\newacronym{mimo}{MIMO}{multiple-input multiple-output}
\newacronym{af}{AF}{amplify-and-forward}
\newacronym{df}{DF}{decode-and-forward}
\newacronym{cf}{CF}{compress-and-forward}
\newacronym{mwrc}{MWRC}{multi-way relay channel}
\newacronym{dmmwrc}{DM-MWRC}{discrete memoryless multi-way relay channel}
\newacronym{pde}{PDE}{partial data exchange}
\newacronym{fde}{FDE}{full data exchange}
\newacronym{iid}{i.i.d.\@}{independent and identically distributed}
\newacronym{di}{DI} {difference of increasing}
\newacronym{dc}{DC}{difference of convex}
\newacronym{mm}{MM}{mixed monotonic}
\newacronym{mmp}{MMP}{mixed monotonic programming}
\newacronym{awgn}{AWGN}{additive white Gaussian noise}
\newacronym{awg}{AWG}{additive white Gaussian}
\newacronym{sic}{SIC}{successive interference cancellation}
\newacronym{snr}{SNR}{signal-to-noise ratio}
\newacronym{sinr}{SINR}{signal to interference plus noise ratio}
\newacronym{inr}{INR}{interference to noise ratio}
\newacronym{zf}{ZF}{zero-forcing}
\newacronym{mrt}{MRT}{maximum ratio transmission}
\newacronym{mmse}{MMSE}{minimum mean square error}
\newacronym{sud}{SUD}{single user decoding}
\newacronym{dof}{DoF}{degrees of freedom}
\newacronym{gdof}{GDoF}{generalized degrees of freedom}
\newacronym{nnc}{NNC}{noisy network coding}
\newacronym{dmn}{DMN}{discrete memoryless network}
\newacronym{csi}{CSI}{channel state information}
\newacronym{pmf}{pmf}{probability mass function}
\newacronym{dmic}{DM-IC}{discrete memoryless interference channel}
\newacronym{ic}{IC}{interference channel}
\newacronym{gic}{GIC}{Gaussian interference channel}
\newacronym{if}{IF}{interference}
\newacronym{ee}{EE}{energy efficiency}
\newacronym{gee}{GEE}{global energy efficiency}
\newacronym{tin}{TIN}{treating interference as noise}
\newacronym{snd}{SND}{simultaneous non-unique decoding}
\newacronym{sd}{SD}{simultaneous decoding}
\newacronym{hk}{HK}{Han-Kobayashi}
\newacronym{rs}{RS}{rate splitting}
\newacronym{rf}{RF}{radio frequency}
\newacronym{pa}{PA}{power amplifier}
\newacronym{lna}{LNA}{low noise amplifier}
\newacronym{lo}{LO}{local oscillator}
\newacronym{adc}{ADC}{analog-to-digital converter}
\newacronym{dac}{DAC}{digital-to-analog converter}
\newacronym{dsp}{DSP}{digital signal processing}
\newacronym{brd}{BRD}{best response dynamics}
\newacronym{br}{BR}{best response}
\newacronym{ne}{NE}{Nash equilibrium}
\newacronym{lhs}{LHS}{left-hand side}
\newacronym{rhs}{RHS}{right-hand side}
\newacronym{ran}{RAN}{radio access network}
\newacronym{qos}{QoS}{Quality of Service}
\newacronym{ngmn}{NGMN}{Next Generation Mobile Networks}
\newacronym{cap}{CAP}{Capacity Adaptation}
\newacronym{bwa}{BW}{Bandwidth Adaptation}
\newacronym{prb}{PRB}{physical resource block}
\newacronym{se}{SE}{spectral efficiency}
\newacronym{tp}{TP}{throughput}
\newacronym{bs}{BS}{base station}
\newacronym{ue}{UE}{user equipment}
\newacronym{mop}{MOP}{multi-objective optimization problem}
\newacronym{gda}{GDA}{generalized Dinkelbach's algorithm}
\newacronym{midcp}{MIDCP}{mixed integer disciplined convex programming}
\newacronym{lp}{LP}{linear program}
\newacronym{brb}{BRB}{branch reduce and bound}
\newacronym{bb}{BB}{branch and bound}
\newacronym{sit}{SIT}{successive incumbent transcending}
\newacronym{oma}{OMA}{orthogonal multiple access}
\newacronym{noma}{NOMA}{non-orthogonal multiple access}
\newacronym{wlog}{w.l.o.g.\@}{without loss of generality}
\newacronym{lsc}{l.s.c.\@}{lower semi-continuous}
\newacronym{usc}{u.s.c.\@}{upper semi-continuous}
\newacronym{kkt}{KKT}{Karush-Kuhn-Tucker}
\newacronym{ptp}{PTP}{point-to-point}

\newacronym{htee}{HTEE}{high throughput energy efficiency}
\glsenableentrycount
\makeglossaries

\usetikzlibrary{positioning}
\usetikzlibrary{calc}
\usetikzlibrary{math}
\usetikzlibrary{fit}
\usetikzlibrary{intersections}
\usetikzlibrary{decorations.pathreplacing}

\usetikzlibrary{external}

\pgfkeys{/pgfplots/.cd,
	hk/.style={blue,mark=*},
	snd/.style={red,mark=square*},
	ian/.style={brown!60!black,mark=triangle*},
}

\makeatletter
\newcommand\transformxdimension[1]{
    \pgfmathparse{((#1/\pgfplots@x@veclength)+\pgfplots@data@scale@trafo@SHIFT@x)/10^\pgfplots@data@scale@trafo@EXPONENT@x}
}
\newcommand\transformydimension[1]{
    \pgfmathparse{((#1/\pgfplots@y@veclength)+\pgfplots@data@scale@trafo@SHIFT@y)/10^\pgfplots@data@scale@trafo@EXPONENT@y}
}
\makeatother

\crefname{equation}{}{}
\crefrangeformat{equation}{(#3#1#4)--(#5#2#6)}

\DeclareMathOperator*{\argmax}{arg\,max}
\DeclareMathOperator*{\argmin}{arg\,min}
\let\card=\abs

\let\vec\bm

\allowdisplaybreaks[1]

\DeclareSIUnit \dBm {dBm}
\DeclareSIUnit \dBW {dBW}
\DeclareSIUnit \bpcu {bpcu}

\usepackage{etoolbox}
\AtBeginEnvironment{equation}{\small}
\AtEndEnvironment{equation}{\normalfont}
\AtBeginEnvironment{gather}{\small}
\AtEndEnvironment{gather}{\normalfont}
\AtBeginEnvironment{split}{\small}
\AtEndEnvironment{split}{\normalfont}
\AtBeginEnvironment{equation*}{\small}
\AtEndEnvironment{equation*}{\normalfont}
\AtBeginEnvironment{align}{\small}
\AtEndEnvironment{align}{\normalfont}
\AtBeginEnvironment{align*}{\small}
\AtEndEnvironment{align*}{\normalfont}
\AtBeginEnvironment{alignat}{\small}
\AtEndEnvironment{alignat}{\normalfont}
\AtBeginEnvironment{flalign}{\small}
\AtEndEnvironment{flalign}{\normalfont}
\AtBeginEnvironment{multline}{\small}
\AtEndEnvironment{multline}{\normalfont}
\AtBeginEnvironment{multline*}{\small}
\AtEndEnvironment{multline*}{\normalfont}

\DeclareFontFamily{U}{mathx}{\hyphenchar\font45}
\DeclareFontShape{U}{mathx}{m}{n}{
      <5> <6> <7> <8> <9> <10>
      <10.95> <12> <14.4> <17.28> <20.74> <24.88>
      mathx10
      }{}
\DeclareSymbolFont{mathx}{U}{mathx}{m}{n}
\DeclareMathSymbol{\bigtimes}{1}{mathx}{"91}

\newtheorem{theorem}{Theorem}

\newtheorem{remark}{Remark}

\newenvironment{optprob}{\begin{equation}\left\{\begin{aligned}}{\end{aligned}\right.\end{equation}\ignorespacesafterend}

\pgfplotscreateplotcyclelist{default}{%
	red,solid,mark=*\\%
	blue,densely dashed,mark=square*\\%
	brown!60!black,densely dashdotted,every mark/.append style={solid},mark=diamond*\\%
}

\pgfplotsset{every axis/.append style = {cycle list name = default}}

\begin{document}
\bstctlcite{IEEEexample:BSTcontrol}
\title{Hierarchical Resource Allocation:\\Balancing Throughput and Energy Efficiency\\in Wireless Systems}

\author{\IEEEauthorblockN{Bho Matthiesen\IEEEauthorrefmark{1}, Eduard A. Jorswieck\IEEEauthorrefmark{2}, and Petar Popovski\IEEEauthorrefmark{3}\IEEEauthorrefmark{1}}
	\IEEEauthorblockA{\IEEEauthorrefmark{1}University of Bremen, Department of Communications Engineering, Germany, email: bho.matthiesen@tu-dresden.de\\\IEEEauthorrefmark{2}TU Braunschweig, Department of Information Theory and Communication Systemes, Germany, email: e.jorswieck@tu-bs.de\\\IEEEauthorrefmark{3}Aalborg University, Department of Electronic Systems, Denmark, email: petarp@es.aau.dk}
\thanks{
The work of B.~Matthiesen and P.~Popovski is supported in part by the German Research Foundation (DFG) through Germany's Excellence Strategy under Grant EXC 2077 (University Allowance).
The work of E.~A.~Jorswieck is supported in part by the
DFG under Grant JO 801/24-1.
}%
}

\maketitle

\begin{abstract}
	A main challenge of 5G and beyond wireless systems is to efficiently utilize the available spectrum and simultaneously reduce the energy consumption.
	From the radio resource allocation perspective, the solution to this problem is to maximize the energy efficiency instead of the throughput. This results in the optimal benefit-cost ratio between data rate and energy consumption. 
	It also often leads to a considerable reduction in throughput and, hence, an underutilization of the available spectrum.
	Contemporary approaches to balance these metrics based on multi-objective programming theory often lack operational meaning and finding the correct operating point requires careful experimentation and calibration. Instead, we
	propose the novel concept of hierarchical resource allocation where conflicting objectives are ordered by their importance. This results in a resource allocation algorithm that strives to minimize the transmit power while keeping the data rate close the maximum achievable throughput. In a typical multi-cell scenario, this strategy is shown to reduces the transmit power consumption by 65\% at the cost of a 5\% decrease in throughput. Moreover, this strategy also saves energy in scenarios where global energy efficiency maximization fails to achieve any gain over throughput maximization.
\end{abstract}
\glsresetall


\section{Motivation and Problem Statement} \label{sec:intro}
The goal of resource allocation in communication networks is to best utilize the available resources ensuring good \cgls{qos} to all users. While the \cgls{qos} constraints are mainly determined by the user's requirements or network slice configuration, the choice of a suitable utility function is entirely up to the operator or system designer \cite{Han2008,Zhang2017,Popovski2018}. Common choices are maximizing the \cgls{tp} to best utilize the available spectrum \cite{Weeraddana2012}, minimizing the total transmit power to save energy \cite{Grandhi1993}, or maximizing the \cgls{ee} to obtain a trade-off between these two \cite{Isheden2012,Zappone2015}. In general, these are conflicting metrics that can not be maximized simultaneously. Indeed, the \cgls{mop}
\begin{equation} \label{eq:mop}
	\max_{\vec p\in\mathcal P} \begin{bmatrix} f_1(\vec p), & f_2(\vec p), & \ldots \end{bmatrix}
\end{equation}
with network utility functions $f_1, f_2, \dots$ is known to posses an infinite number of noninferior solutions \cite{Zadeh1963}. The \cgls{mop} \cref{eq:mop} is usually solved by transforming it into a scalar optimization problem, e.g., with the scalarization approach \cite{Zhang2010a} where the weighted sum of the objectives is maximized, i.e.,
\begin{equation*}
	\max_{\vec p\in\mathcal P} \sum\nolimits_{i} w_i f_i(\vec p),
\end{equation*}
or by the utility profile approach \cite{Zhang2010a} where the intersection of a ray in the direction $\vec w$ and the outer boundary of the performance region is computed, i.e.,
\begin{equation*}
	\max_{t, \vec p\in\mathcal P}\enskip t \quad\mathrm{s.\,t.}\quad \forall i: t w_i \le f_i(\vec p).
\end{equation*}
Both methods obtain Pareto optimal points but share the weakness that
the weights $\vec w$ often have no operational meaning and need to be chosen heuristically or by experimentation.

For example, consider balancing the \cgls{tp} with the total transmit power. This problem is formally stated as
\begin{optprob}
	& \underset{\vec p, \vec r}{\mathrm{max}}
	&& \begin{bmatrix} \sum_i r_i, & -\sum_i p_i \end{bmatrix} \\
	& \mathrm{s.\,t.}
	&& \vec r \in \mathcal R(\vec p) \cap \mathcal Q,\quad
	\vec 0 \le \vec p \le \vec P
	\label{opt:ramop}
\end{optprob}
where $\vec P$ is the maximum transmit power, $\mathcal R(\vec p)$ the achievable rate region, and $\mathcal Q$ contains the \cgls{qos} constraints. After scalarization, the problem becomes
\begin{optprob} \label{opt:ramopscal}
	& \underset{\vec p, \vec r}{\mathrm{max}}
	&& w_1 \sum_i r_i - w_2 \sum_i p_i \\
	& \mathrm{s.\,t.}
	&& \vec r \in \mathcal R(\vec p) \cap \mathcal Q,\quad
	\vec 0 \le \vec p \le \vec P
\end{optprob}
with nonnegative weights $w_1$, $w_2$. By varying these weights such that $w_1 + w_2 = 1$, the convex hull of the Pareto boundary is obtained. However, these weights do not have much operational meaning and there is no other guidance than experience or experimentation to choose them for a given system.
Another approach to balance \cgls{tp} and transmit power is the notion of \cgls{gee}, which is defined as the benefit-cost ratio of system throughput and total dissipated power, i.e.,
$\mathrm{GEE} = \frac{\sum_i r_i}{\sum_i \mu_i p_i + P_c}$,
where $\mu_i \ge 0$ and $P_c > 0$ are modeling constants reflecting the power amplifier inefficiency and static circuit power consumptions.
Maximizing the \cgls{gee} results in a Pareto optimal solution of \cref{opt:ramop} \cite[p.~241]{Zappone2015} and has a well defined operational meaning.
With energy and spectrum being similarly scarce resources, the 
\cgls{tp} and \cgls{gee} are considered to be the most important network utility functions in 5G and beyond networks.

\begin{figure}
\centering
\pgfplotsset{every axis/.append style={
		thick,
		ylabel near ticks,
		xlabel near ticks,
		grid=none,
		no markers,
		xmin = -12,
		xmax = 30,
		xtick=\empty,
		ytick=\empty,
		xlabel={\footnotesize Power},
		width = .6*\axisdefaultwidth,
		height = 80pt,
}}
{\tikzexternaldisable\ref{fig:tpee:legend}}\\[1.5pt]
\begin{tikzpicture}[baseline]
	\begin{axis} [
			legend style={font=\footnotesize,thick,/tikz/column 2/.style={column sep=10pt}, inner ysep = 1pt, inner xsep=10pt},
			legend to name = fig:tpee:legend,
			legend columns = 2,
			legend cell align=left,
		]

		\pgfplotstableread[col sep=comma]{throughput.dat}\tbl

		\addplot+[smooth] table[y=TIN] {\tbl};
		\addlegendentry{TP max};

		\addplot+[smooth] table[y=GEE] {\tbl};
		\addlegendentry{GEE max};

		\node [anchor=north west] at (rel axis cs:0,1) {TP};
	\end{axis}
\end{tikzpicture}
\begin{tikzpicture}[baseline]
	\begin{axis} [
		]

		\pgfplotstableread[col sep=comma]{gee.dat}\tbl

		\addplot+[smooth] table[y=TIN] {\tbl};

		\addplot+[smooth] table[y=GEE] {\tbl};

		\node [anchor=north west] at (rel axis cs:0,1) {GEE};
	\end{axis}
\end{tikzpicture}
\caption{Typical solution of \cgls{tp} and \cgls{gee} maximization.}
\label{fig:tpee}
\vspace{-2ex}
\end{figure}
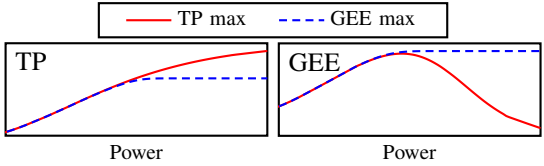

A qualitative solution of \cgls{tp} and \cgls{gee} maximization in wireless interference networks is displayed in \cref{fig:tpee}. While leading to similar operating points in the low \cgls{snr} regime, it is characteristic for the \cgls{gee} to saturate. The link and power budget in a wireless network often allow for an operating point far in this saturation region. In such a scenario, selecting the operating point by \cgls{tp} or \cgls{gee} maximization either results in poor \cgls{ee} or in low spectral efficiency.
Thus, it has been proposed in \cite{Aydin2017} to balance \cgls{tp} and \cgls{gee} with multi-objective programming theory. While the obtained performance region provides valuable insights for system design, the weights still have little operational meaning.
A more straightforward method is to maximize the \cgls{gee} under \cgls{qos} constraints which is expected to provide the best rate-energy trade-off while still providing satisfactory service to all users.

\tikzsetnextfilename{TPEE}
\tikzpicturedependsonfile{throughput.dat}
\tikzpicturedependsonfile{gee.dat}

Taking the operator's perspective, saving energy is just a secondary concern, while generating revenue from their costly equipment and spectrum licenses is the primary goal. This requires good service quality to outperform competitors and thereby ensure customer loyalty. Satisfying \cgls{qos} constraints and providing good connectivity is undoubtedly the foundation for good service but from there it's up to the operator to choose an operating point in the resource allocation design space. A viable strategy is to prioritize high service quality and minimize energy consumption as a secondary objective to reduce operational expenditures and further increase revenue. This could be achieved by solving \cref{opt:ramopscal} with $w_1 \gg w_2$. A more rigorous approach is to use 
lexicographic ordering \cite[\S 4.2]{Miettinen1999}, a recursive multi-objective programming technique where objectives are strictly ordered by priority. In the context of this paper and \cref{opt:ramop}, a lexicographic ordering approach is to maximize the \cgls{tp} first and then select the solution with lowest transmit power, i.e.,
$\vec p^\star = \min\{ \sum_i p_i \,|\, \vec p \in\mathcal T^\star \}$
where $\mathcal T^\star$ is the set of throughput optimal power allocations, i.e.,
$\mathcal T^\star = \argmax\{ \sum r_i \,|\, \vec r \in\mathcal R(\vec p),\ \vec 0\le\vec p\le\vec P\}$.
When the solution to the \cgls{tp} maximization problem is (almost) unique, i.e., the volume of $\mathcal T^\star$ is close to zero, the possible power reduction due to this approach is negligible. 
However, significant gains are possible by slightly relaxing this strict ordering of the objectives.
For example, the 
goal could be to achieve at least \SI{95}{\percent} of the maximum \cgls{tp} instead of strictly maximizing it, i.e.,
$\vec p^\star = \min\{ \sum_i p_i \,|\, \sum_i r_i \ge 0.95 \cdot r_{\Sigma}^\star,\ \vec r\in\mathcal R(\vec p) \}$, where $r_{\Sigma}^\star$ is the optimal value of the \cgls{tp} maximization problem.
Selecting a power allocation within this tight \cgls{tp} region leaves more freedom than lexicographic ordering, while still ensuring high service quality.

Leaving economical considerations aside, there are plenty of other technical motivations to strictly prioritize a high \cgls{tp} over other metrics.
One application arises from cross-layer optimization where the queue of a \cgls{bs} needs to be stabilized. Regardless of the underlying queuing model, 
the total storage capacity is essentially limited by the \gls{bs}'s installed memory. The \cgls{tp} determines the maximum departure rate of this joint queue and, hence, maximizing the \cgls{tp} ultimately enlarges the stability region.
Please refer to \cite{Weeraddana2012} for further application examples.

The goal of this paper is to obtain a hierarchical Pareto optimal solution of \cref{opt:ramop} for wireless interference networks,
and to evaluate the benefits of this approach over \cgls{gee} maximization numerically. As the resulting optimization problem is NP-hard and numerically very challenging, this requires the careful design of a solution algorithm. We show that, by reducing the \cgls{tp} by just \SI{5}{\percent}, almost \SI{65}{\percent} of transmit power can be saved in a typical wireless network.

\subsection{System Model}
We consider a Gaussian interference network with power allocation $\vec p = (p_1, p_2, \dots)$ and average power constraint $\vec P$. The receive \gls{sinr} is
$\frac{\alpha_i p_i}{\sum_{j\neq i} \beta_{i j} p_j + \sigma_i^2}$
and, under the assumption that interference is treated as noise, asymptotic error free communication is possible at all rates $\vec r$ satisfying
\begin{equation*}
	r_i \le B \log\left( 1 + \frac{\alpha_i p_i}{\sum_{j\neq i} \beta_{i j} p_j + \sigma_i^2} \right)
\end{equation*}
for all $i$, where $B$ is the communication bandwidth.
In this setting, $\alpha_i$ is the effective channel gain of the direct channel from transmitter $i$ to receiver $i$, $\beta_{i j}$ are the effective channels from transmitter $j$ to receiver $i$, and $\sigma_i^2$ is the variance of circularly-symmetric complex Gaussian noise.

This adequately models the effective channel for multi-antenna transmission in 5G networks after precoder matrix selection \cite[\S 11]{Dahlman2018}, for multi-cell networks with overlapping frequencies, and for dense \gls{leo} satellite constellations \cite{satwp}. Other applications include, e.g., massive MIMO and relay-assisted CoMP networks \cite{Zappone2016}.

\section{Hierarchical Optimization}
Hierarchical optimization \cite[\S 4.2.2]{Miettinen1999}, \cite{Bestle1997} is a solution method for the \cgls{mop} \cref{eq:mop} where the objectives are arranged a~priori by their absolute importance. \cGls{wlog}, assume that $f_i$ is more important to the system designer than $f_{i+1}$. The optimization is carried out recursively by first maximizing $f_1$ and ignoring all other objectives $f_2, f_3, \dots$. Then, the next objective $f_2$ is maximized with additional constraint that the value of $f_1$ is close to the optimal value of the previous optimization. Mathematically, the $i$th optimization problem is
\begin{equation*}
	\max_{\vec p\in\mathcal D_i} f_i(\vec x)
	\quad\text{with}\quad
	\mathcal D_i = \{ \vec p \in\mathcal D_{i-1} \,|\, f_{i-1}(\vec p) \ge \omega_{i-1} f_{i-1}^\star \}
\end{equation*}
for all $i > 1$ and some initial feasible set $\mathcal D_1$. Here, $f_{i}^\star$ denotes the optimal value of the $i$th problem and $\omega_1, \omega_2, \dots$ are so-called worsening factors. These are selected a~priori by the system designer and have, contrary to the weights in the multi-objective programming solution approaches discussed in \cref{sec:intro}, a clearly defined operational meaning in many engineering problems. Lexicographic ordering \cite[\S 4.2]{Miettinen1999} is a special case of this approach obtained by setting all worsening factors to one.
For a \cgls{mop} with two objectives, the second (and final) optimization step is equivalent to the $\varepsilon$-constraint method \cite[\S 3.2]{Miettinen1999} and its solution is a strictly Pareto optimal point if it is unique \cite[Thm.~3.2.4]{Miettinen1999}.

Applying this approach to the \cgls{mop}~\cref{opt:ramop} and prioritizing the \cgls{tp} over the transmit power, we obtain two scalar optimization problems\footnote{The constant $B$ is inessential and moved into $r_{i,\mathrm{min}}$ for notational clarity.}
\begin{subequations} \label{opt:rsum}
\begin{empheq}[left=\empheqlbrace]{alignat=2}
	& \underset{\vec p, \vec r}{\mathrm{max}}
	&\quad& \sum\nolimits_i \log\left( 1 + \frac{\alpha_i p_i}{\sum_{j\neq i} \beta_{i j} p_j + \sigma_i^2} \right) \\
	& \mathrm{s.\,t.}
	&& \forall i: \log\left( 1 + \frac{\alpha_i p_i}{\sum_{j\neq i} \beta_{i j} p_j + \sigma_i^2} \right) \ge r_{i,\mathrm{min}} \label{opt:rsum:qos} \\
	&&&	\vec 0 \le \vec p \le \vec P
\end{empheq}
\end{subequations}
for minimum rate constraints $r_{i,\mathrm{min}} \ge 0$, and
\begin{subequations} \label{opt:pmin}
\begin{empheq}[left=\empheqlbrace]{alignat=2}
	& \underset{\vec p, \vec r}{\mathrm{min}}
	&\quad& \sum\nolimits_i p_i \\
	& \mathrm{s.\,t.}
	&& \sum\nolimits_i \log\left( 1 + \frac{\alpha_i p_i}{\sum_{j\neq i} \beta_{i j} p_j + \sigma_i^2} \right) \ge \omega r_\Sigma^\star \label{opt:pmin:rsum} \\
	&&& \forall i: \log\left( 1 + \frac{\alpha_i p_i}{\sum_{j\neq i} \beta_{i j} p_j + \sigma_i^2} \right) \ge r_{i,\mathrm{min}} \label{opt:pmin:qos}\\
	&&&	\vec 0 \le \vec p \le \vec P
\end{empheq}
\end{subequations}
where $r_{\Sigma}^\star$ is the optimal value of \cref{opt:rsum} and $\omega\in[0, 1]$ is the worsening factor that determines the acceptable \cgls{tp} reduction. Clearly, it is necessary to solve \cref{opt:rsum} before \cref{opt:pmin}.

Both problems \cref{opt:rsum,opt:pmin} are challenging global optimization problems due to the nonconvexity of the objective in \cref{opt:rsum} and constraint \cref{opt:pmin:rsum}. In particular, \cref{opt:rsum} is known to be NP-hard \cite{Luo2008}, and, hence, \cref{opt:pmin} is also NP-hard due to constraint \cref{opt:pmin:rsum}. While \cref{opt:rsum} can be solved efficiently using the \cgls{mmp} framework as discussed next, problem~\cref{opt:pmin} needs a novel algorithm that is developed in \cref{sec:sit}.

\subsection{Solution of Problem~\eqref{opt:rsum}}
\label{sec:mmp}
\cGls{mmp} is a global optimization framework that exploits partial monotonicity in the objective and constraints \cite{mmp}. It is much more versatile than classical monotonic optimization \cite{Tuy2000}  and shows tremendous performance gains over state-of-the-art algorithms for global optimal power allocation in interference networks and other scenarios \cite[\S IV]{mmp}.

The concept of \cgls{mm} functions generalizes differences of increasing functions. Let $\mathcal M_0$ be a box in $\mathds R^n$, i.e.,
$\mathcal M_0 = [\vec r^0, \vec s^0] = \{ \vec x\in\mathds R^n \,|\, \forall i: r_i^0 \le x_i \le s_i^0 \}$.
A continuous function  $F : \mathds R^n \times \mathds R^n \rightarrow \mathds R$ is called \cgls{mm} function if it satisfies
\begin{align*}
	F(\vec x, \vec y) &\le F(\vec x', \vec y)\qquad \text{if}\ \vec x\le\vec x', \\
	F(\vec x, \vec y) &\ge F(\vec x, \vec y')\qquad \text{if}\ \vec y\le\vec y'. 
\end{align*}
for all $\vec x, \vec x', \vec y, \vec y' \in \mathcal M_0$ and a continuous optimization problem
$\max_{\vec x \in\mathcal D} f(\vec x)$
with compact feasible set $\mathcal D \subseteq \mathds R^n$ is called \cgls{mmp} problem if there exists an \cgls{mm} function $F$ such that
$F(\vec x, \vec x) = f(\vec x)$
for all $\vec x \in \mathcal M_0$, where $\mathcal M_0 \supseteq \mathcal D$ encloses $\mathcal D$.
The \cgls{mmp} framework \cite{mmp} solves such a problem very efficiently with global optimality using a \cgls{bb} procedure.

Applying the \cgls{mmp} framework requires \cgls{mm} representations of the objective and constraint functions in \cref{opt:rsum}. For the objective, such a function is $\vec x, \vec y \mapsto \sum_i R_i(\vec x, \vec y)$ with \cite[\S IV-A]{mmp}
\begin{equation} \label{eq:mmri}
	R_i(\vec x, \vec y) = \log\left( 1 + \frac{\alpha_i x_i}{\sum_{j\neq i} \beta_{i j} y_j + \sigma_i^2} \right).
\end{equation}
Likewise, the \cgls{qos} constraints have \cgls{mm} representation $\vec x, \vec y \mapsto r_{i,\min} - R_i(\vec x, \vec y)$. Theoretically, such \cgls{mm} constraints lead to an algorithm without guaranteed finite convergence. This is, because for general \cgls{mm} constraints and some boxes $\mathcal M$, it is impossible to determine whether $\mathcal M\cap\mathcal D$ contains feasible points or not \cite[\S III-A]{mmp}. However, in practise this is seldom a problem for typical minimum rate constraints as in \cref{opt:rsum:qos}.

The \cgls{mmp} framework is also applicable to \cref{opt:pmin}. However, the minimum sum rate constraint in \cref{opt:pmin:rsum} is very tight and leads to a tiny feasible set compared to $\mathcal M^0 = [\vec 0, \vec P]$. This results in impractically slow convergence of the \cgls{mmp} procedure.
In the next section, we develop an algorithm with much faster and provably finite convergence.

\section{Successive Incumbent Transcending Scheme} \label{sec:sit}
The main challenge in solving \cref{opt:pmin} with the \cgls{mmp} framework is constraint \cref{opt:pmin:rsum}. An efficient solution to this problem is the \cgls{sit} scheme developed in \cite{Tuy2009}. The main idea is to
solve a sequence of easily implementable feasibility problems. Specifically, given
a real number $\gamma$, the core problem of the \cgls{sit} algorithm is to check whether \cref{opt:pmin}
has a feasible solution $\vec p$ satisfying $\sum_i p_i \le \gamma$, or, else, establish that
no such $\vec p$ exists. In this manner, a sequence of
feasible points (``incumbents'') with decreasing objective value is generated until
no point with lesser objective value than the current best solution $\gamma$ exists.

Consider the optimization problem
\begin{equation}
	\min_{\vec x\in\mathcal M_0}\enskip f(\vec x) \quad\mathrm{s.\,t.}\quad g(\vec x) \le 0
	\label{opt:sit:primal}
\end{equation}
which generalizes \cref{opt:pmin} and
assume that $f$ is a nondecreasing function, $g$ has an \cgls{mm} representation, and $\mathcal M_0$ is a box. The outlined \cgls{sit} scheme for this problem is given in \cref{alg:sit}.

\begin{algorithm}
	\def\tuyref{\cite[Sect.~7.5.1]{Tuy2016}}
	\caption{\cGls{sit} Scheme \tuyref}\label{alg:sit}
	\small
	\centering
	\begin{minipage}{\linewidth-1em}
		\begin{enumerate}[label=\textbf{Step \arabic*},ref=Step~\arabic*,start=0,leftmargin=*]
			\item \label{alg:sit:step0} Initialize $\bar{\vec x}$ with the best known feasible solution and set $\gamma = f(\bar{\vec x}) - \eta$; otherwise do not set $\bar{\vec x}$ and choose  some $\gamma \le f(\vec x)$ $\forall \vec x \in \mathcal M_0 : g(\vec x) \le 0$.
			\item\label{alg:sit:step1} Check if \cref{opt:pmin} has a feasible solution $\vec x$ satisfying $f(\vec x) \ge \gamma$; otherwise, establish that no such feasible $\vec x$ exists and go to \ref{alg:sit:step3}.
			\item Update $\bar{\vec x} \gets \vec x$ and $\gamma \gets f(\bar{\vec x}) - \eta$. Go to \ref{alg:sit:step1}.
			\item\label{alg:sit:step3} Terminate: If  $\bar{\vec x}$ is set, it is an $\eta$-optimal solution; else Problem \cref{opt:pmin} is infeasible.
		\end{enumerate}
	\end{minipage}
\end{algorithm}

Implementing the feasibility check in \ref{alg:sit:step1} of \cref{alg:sit} efficiently is crucial.
Consider the optimization problem
\begin{equation}
	\min_{\vec x\in\mathcal M_0}\enskip g(\vec x) \quad\mathrm{s.\,t.}\quad  f(\vec x) \le \gamma
	\label{opt:sit:dual}
\end{equation}
which is
dual to \cref{opt:sit:primal} in the sense that if the optimal value of \cref{opt:sit:dual} is greater than zero, the optimal value of \cref{opt:sit:primal} is greater than $\gamma$ \cite[Prop.~7.13]{Tuy2016}.
Thus, any point $\vec x'$ in the feasible set of \cref{opt:sit:dual} with objective value less than zero is also a feasible point in \cref{opt:sit:primal} with objective value less than $\gamma$. We can solve \cref{opt:sit:primal} sequentially by solving \cref{opt:sit:dual} with a \cgls{bb} method.

At first, this approach seems to increase the computational complexity significantly because if \cref{opt:sit:primal} is nonconvex, then so is \cref{opt:sit:dual}. However, given that $f$ has favorable properties,\footnote{Such favorable properties could be, e.g., linearity, convexity, or being increasing.} problem~\cref{opt:sit:dual} might be considerably easier to solve than \cref{opt:sit:primal}. Moreover, the \cgls{sit} scheme can be combined with the \cgls{bb} procedure that solves \cref{opt:sit:dual}. This eliminates the need to solve \cref{opt:sit:dual} multiple times.

Exploiting the properties of \cgls{mm} functions,
we can obtain a lower bound on the objective value of \cref{opt:sit:dual} over a box $\mathcal M = [\vec r, \vec s]$ from its \cgls{mm} representation $G$ as
\begin{equation*}
	\min_{\vec x\in\mathcal M: f(\vec x) \le \gamma} g(\vec x) \ge \min_{\vec x\in\mathcal M} G(\vec x, \vec x) \ge \min_{\vec x, \vec y\in\mathcal M} G(\vec x, \vec y) = G(\vec r, \vec s).
\end{equation*}
Together with an exhaustive rectangular subdivision \cite{Tuy2016}, this bound leads to a convergent \cgls{bb} procedure that can be incorporated into the \cgls{sit} scheme.

The complete algorithm is stated in \cref{alg:sitbb}. It involves a parameter $\varepsilon$ that is related to the concept of $\varepsilon$-essential feasibility explained in \cite{tsp2018}. Its primary roles are to exclude numerically instable points from the feasible set and ensure finite convergence of the algorithm. The latter is established in the theorem below. This is the first algorithm that combines the \cgls{mmp} approach with the \cgls{sit} scheme.
\begin{theorem}
	\Cref{alg:sitbb} converges in finitely many steps to the $(\varepsilon, \eta)$-optimal solution of \cref{opt:sit:primal} or establishes that no such solution exists.
\end{theorem}
\begin{IEEEproof}[Proof sketch]
	By virtue of \cite[Prop.~7.14]{Tuy2016} a \cgls{bb} procedure for solving \cref{opt:sit:dual} with pruning criterion $G(\vec r, \vec s) > -\varepsilon$ and stopping criterion $g(\vec r) < 0$ or $\mathscr R_k = \emptyset$ implements \ref{alg:sit:step1} in \cref{alg:sit}. Thus, start with the \cgls{mmp} algorithm in \cite[Alg.~1]{mmp} for \cref{opt:sit:dual} and modify it according to the previous sentence. Establishing finite convergence is a minor modification of \cite[Thm.~1]{mmp}.
	Next, integrate the \cgls{sit} scheme in \cref{alg:sit} into this procedure: move the termination criterion $g(\vec r) < 0$ into the incumbent update in \ref{alg:sitbb:incumbent} and update $\gamma_k$ if a box satisfies this criterion. It remains to show that continuing the procedure after updating $\gamma_k$ preserves convergence. This part of the proof follows along the lines of the proof of \cite[Thm.~1]{tsp2018}.
\end{IEEEproof}

\begin{algorithm}
	\caption{\cGls{sit} Algorithm for \cref{opt:sit:primal}}\label{alg:sitbb}
	\small
	\centering
	\begin{minipage}{\linewidth-1em}
	\begin{enumerate}[label=\textbf{Step \arabic*},ref=Step~\arabic*,start=0,leftmargin=*]
		\item\label{alg:sitbb:init} {\bfseries (Initialization)} Set $\varepsilon, \eta > 0$, 
			Let $k=1$ and $\mathscr R_0 = \{ \mathcal M_0 \}$.
			If available, initialize $\bar{\vec x}^0$ with the best known feasible solution and set $\gamma_k = f(\bar{\vec x}) - \eta$. Otherwise, do not set $\bar{\vec x}^0$ and choose $\gamma \ge f(\vec x)$ for all feasible $\vec x$.
		\item\label{alg:sitbb:branch} {\bfseries (Branching)} Let $\mathcal M_k = [\vec r^k, \vec s^k]$ be the oldest box in $\mathscr R_{k-1}$. Bisect $\mathcal M_k$ via $(\vec v^k, j_k)$ with $j_k \in \argmax_j s^k_j - r^k_j$ and $\vec v^k = \frac{1}{2} (\vec s^k + \vec r^k)$, i.e., compute
				\begin{equation*}
				\begin{aligned}
					\mathcal M^- &= \{ \vec x  \,|\, r_j^k \le x_j \le v_j^k,\ r_i^k \le x_i \le s_i^k\ (i\neq j) \} \\
					\mathcal M^+ &= \{ \vec x \,|\, v_j^k \le x_j \le s_j^k,\ r_i^k \le x_i \le s_i^k\ (i\neq j) \},
				\end{aligned}
				\end{equation*}
			and set $\mathscr P_{k} = \{\mathcal M^k_-, \mathcal M^k_+\}$.
		\item\label{alg:sitbb:reduction} {\bfseries (Reduction)} Replace each box in $\mathcal M\in\mathscr P_k$ with some $\mathcal M'$ such that $\mathcal M'\subseteq\mathcal M$ and
				\begin{multline}
					\min\{ g(\vec x) \,|\, f(\vec x) \le \gamma_k,\ \vec x\in\mathcal M \}
					\\=
					\min\{ g(\vec x) \,|\, f(\vec x) \le \gamma_k,\ \vec x\in\mathcal M' \}
				\end{multline}

			\item\label{alg:sitbb:incumbent} {\bfseries (Incumbent)} Let $\mathscr I = \{ \vec r \,|\, [\vec r, \vec s]\in\mathscr P_k,\ g(\vec r) \le 0 \}$. If not empty, set $\vec r^k = \argmin_{\vec r\in\mathscr I} f(\vec r)$. If $\bar{\vec x}^{k-1}$ is not set or $f(\vec r^k) < \gamma_{k-1}+\eta$, set $\bar{\vec x} = \vec r^k$ and $\gamma_k = f(\vec r^k) - \eta$. In all other cases, set $\bar{\vec x}^k = \bar{\vec x}^{k-1}$ and $\gamma_k = \gamma_{k-1}$.
		\item\label{alg:sitbb:prune} {\bfseries (Pruning)} Delete every $[\vec r, \vec s]\in\mathscr P_k$ with $f(\vec r) \ge \gamma_k$ or $G(\vec r, \vec s) > -\varepsilon$. Let $\mathscr P_k'$ be the collection of remaining sets and set $\mathscr R_k = \mathscr P_k'\cup(\mathscr R_{k-1}\setminus\{\mathcal M_k\})$.
		\item\label{alg:sitbb:terminate} {\bfseries (Termination)} Terminate if $\mathscr R = \emptyset$: If $\bar{\vec x}^k$ is not set, then \cref{opt:sit:primal} is $\varepsilon$-essential infeasible; else $\bar{\vec x}^k$ is an essential $(\varepsilon, \eta)$-optimal solution of \cref{opt:sit:primal}. Otherwise, update $k\gets k+1$ and return to \ref{alg:sitbb:branch}.
	\end{enumerate}
	\end{minipage}
\end{algorithm}

The purpose of the reduction in \ref{alg:sitbb:reduction} is to speed up the convergence. This is achieved by
replacing the box under consideration by a smaller one that still contains all candidate solutions and, thereby,
improves the quality of the computed bounds. One approach to determine this procedure for \cref{alg:sitbb} is to replace $\mathcal M$ by $\mathcal M' = [\vec r', \vec s']$ with
\begin{align} \label{sit:red:lbub}
	r'_i &= \min_{\vec x\in\mathcal M : f(\vec x) \le \gamma_k} x_i,
	&
	s'_i &= \max_{\vec x\in\mathcal M : f(\vec x) \le \gamma_k} x_i
\end{align}
for all $i$. For $f$ nondecreasing, the solution to the first problem is always $r_i$ unless it is infeasible. For the upper bound in \cref{sit:red:lbub}, recall that $\vec r$ minimizes $f(\vec x)$ over $\mathcal M$. Thus, the optimal solution to this optimization problem is to set $x_j = r_j$ for all $j \neq i$. Then, the optimal $x_i = \min\{ \tilde x_i, s_i\}$ where $\tilde x_i$ satisfies
\begin{equation} \label{sit:red:xi}
	f(\vec r + (\tilde x_i - r_i) \vec e_i) = \gamma_k.
\end{equation}

\begin{remark}[Branch selection]
	Most \cgls{bb} procedures select the box with the largest bound for further partitioning. The rationale is that this choice leads to fastest convergence. In practice, when the number of boxes in $\mathcal R_k$ grows very large, this selection rule might become the performance and memory bottleneck of the algorithm. First, it tends to store suboptimal boxes longer than necessary and therefore increases memory consumption. Second, inserting new boxes into $\mathcal R_k$ has complexity $O(\log \card{\mathcal R_k})$. Instead, with the \emph{oldest-first} rule employed in \cref{alg:sitbb} inserting new boxes has constant complexity. Also, every box is visited after a fixed amount of time and, thus, likely to be pruned much earlier than with the best-first rule \cite{mmp}. Since \cref{alg:sitbb} is essentially memory limited, the oldest-first rule performs much better than the standard best-first rule.
\end{remark}

\begin{remark}[Other \cgls{sit} applications]
	Despite its tremendous numerical advantages, the \cgls{sit} approach is currently not widely used. Besides the applications to DC and monotonic optimization problems in \cite{Tuy2009,Tuy2016}, it is only employed in \cite{tsp2018} where it is applied to resource allocation problems with fractional objectives and partial convexity. The implementation most closely related to \cref{alg:sitbb} is the monotonic optimization variant in \cite[\S 11.3]{Tuy2016}. The key advantage of \cref{alg:sitbb} over this procedure is that cumbersome transformations and an auxiliary variable are required to bring \cref{opt:pmin} into a suitable form for \cite[\S 11.3]{Tuy2016}. This leads to much slower convergence due to the extra variable and much looser bounds on the constraints.
\end{remark}

\subsection{Solution of Problem~\eqref{opt:pmin}}
Identify $\mathcal M_0 = [\vec 0, \vec P]$ and $f(\vec p) = \sum_i p_i$. Note that $f(\vec p)$ is an increasing function. \cGls{mm} representations of \cref{opt:pmin:rsum} and \cref{opt:pmin:qos} are
$\vec x, \vec y \mapsto \omega r_\Sigma^\star - \sum_i R_i(\vec y, \vec x)$
and
$\forall i: \vec x, \vec y \mapsto r_{i,\mathrm{min}} - R_i(\vec y, \vec x)$,
respectively,
with $R_i(\vec x, \vec y)$ as in \cref{eq:mmri}. 
They can be merged into a single inequality constraint $\max_i g_i(\vec x) \le 0$ with \cgls{mm} representation
\begin{equation*}
	G(\vec x, \vec y) = \max\Big\{ \omega r_\Sigma^\star - \sum_i R_i(\vec y, \vec x),\ \max_i\big\{ r_{i,\mathrm{min}} - R_i(\vec y, \vec x) \big\}\Big\}
\end{equation*}
due to \cite[Eq.~(9)]{mmp}.
In the reduction step, the solution to \cref{sit:red:xi} is $\tilde x_i = \gamma_k - \sum_{j\neq i} r_j$. Thus, every box $\mathcal M=[\vec r, \vec s]$ in \ref{alg:sitbb:reduction} can be replaced by $[\vec r, \vec s']$ with
$s_i' = \min\{ s_i,\ \gamma_k - \sum\nolimits_{j\neq i} r_j \}$.
With these choices, \cref{alg:sitbb} solves \cref{opt:pmin} in a finite number of iterations.

\section{Numerical Evaluation}

We consider uplink transmission in a \cgls{siso} multi-cell system. \cGlspl{ue} are placed randomly in a rectangular area with edge length \SI{1}{\km}. This area is divided into four equal sized cells with \cGlspl{bs} located at the center of their cell. Path-loss is modeled according to the Hata-COST231 \cite{HataCOST231,Rappaport2002} urban scenario with carrier frequency \SI{1.9}{\giga\Hz}, \SI{30}{\meter} \cgls{bs} height and \SI{8}{\dB} log-normal shadow fading. Small scale effects are modeled as Rayleigh fading. Each \cgls{ue} is associated to the \cgls{bs} with the best channel. Scenarios where more than one \cgls{ue} is associated to a \cgls{bs} are dropped. The receivers have noise spectral density $N_0 = \SI{-174}{\dBm}$ and noise figure $F = \SI{3}{\dB}$.
The communication bandwidth is $B = \SI{180}{\kilo\Hz}$ and the noise power is calculated as $\sigma_i^2 = N_0 F B$. The \cglspl{ue} RF chains have a static power consumption $P_c = \SI{400}{\milli\W}$ and \cglspl{pa} with an efficiency of \SI{25}{\percent}.
No cooperation between \cglspl{bs} is assumed, i.e., interference from other cells is treated as noise. 

\cGls{tp} and \cgls{gee} are maximized using the \cgls{mmp} framework \cite{mmp}.
\Cref{alg:sitbb} is used to solve \cref{opt:pmin} for $\omega = 0.95$, i.e., the obtained resource allocation uses the minimum total transmit power under the constraint that the system \cgls{tp} is not less than \SI{95}{\percent} of the maximum achievable system \cgls{tp}.
We call this resource allocation \gls{htee} for
reasons that will become apparent below. All algorithms obtain the global optimal solution within an absolute tolerance of $\eta = 0.01$. In \cref{alg:sitbb}, we set $\varepsilon = 10^{-5}$. All results are averaged over 1000 \cgls{iid} channel realizations.

\begin{figure}
\centering
\begin{tikzpicture}
	\begin{axis} [
			thick,
			ylabel near ticks,
			xlabel near ticks,
			grid=both,
			minor x tick num = 1,
			minor y tick num = 1,
			no markers,
			legend style={font=\small},
			legend pos=south east,
			legend cell align=left,
			xlabel={Maximum Tx Power P [dBm]},
			ylabel={Throughput [Mbit/s]},
			xmin = -20,
			xmax = 30,
			width=\axisdefaultwidth,
			height=.75*\axisdefaultheight,
		]

		\pgfplotstableread[col sep=comma]{throughput.dat}\tbl

		\addplot+[smooth] table[y=TIN] {\tbl};
		\addlegendentry{TP};

		\addplot+[smooth] table[y=HTEE95] {\tbl};
		\addlegendentry{HTEE \SI{95}{\percent}};

		\addplot+[smooth] table[y=GEE] {\tbl};
		\addlegendentry{GEE};
	\end{axis}
\end{tikzpicture}
\vspace{-2ex}
\caption{Achievable throughput with different resource allocation approaches.}
\label{fig:tp}
\end{figure}
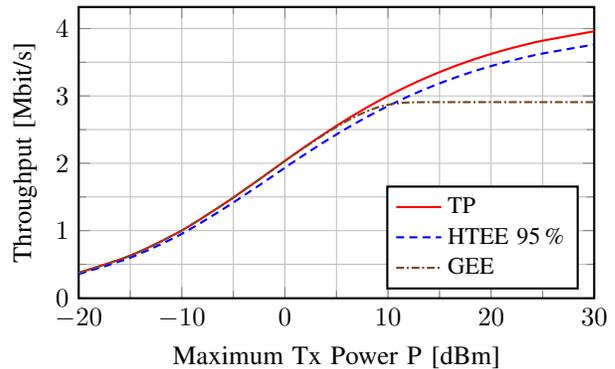

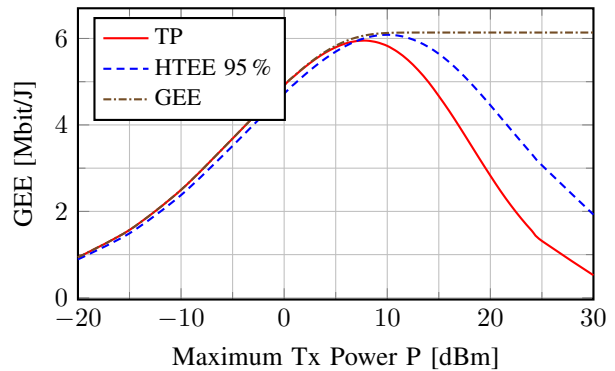
\begin{figure}
\centering
\begin{tikzpicture}
	\begin{axis} [
			thick,
			ylabel near ticks,
			xlabel near ticks,
			grid=both,
			minor x tick num = 1,
			minor y tick num = 1,
			no markers,
			legend style={font=\small},
			legend pos=north west,
			legend cell align=left,
			xlabel={Maximum Tx Power P [dBm]},
			ylabel={\gls{gee} [Mbit/J]},
			xmin = -20,
			xmax = 30,
			width=\axisdefaultwidth,
			height=.75*\axisdefaultheight,
		]

		\pgfplotstableread[col sep=comma]{gee.dat}\tbl

		\addplot+[smooth] table[y=TIN] {\tbl};
		\addlegendentry{TP};

		\addplot+[smooth] table[y=HTEE95] {\tbl};
		\addlegendentry{HTEE \SI{95}{\percent}};

		\addplot+[smooth] table[y=GEE] {\tbl};
		\addlegendentry{GEE};
	\end{axis}
\end{tikzpicture}
\vspace{-2ex}
\caption{Global energy efficiency of the discussed resource allocation strategies.}
\label{fig:gee}
\vspace{-2ex}
\end{figure}

\Cref{fig:tp,fig:gee} display \cgls{tp} and \cgls{gee}, respectively, with very typical behavior. With increasing transmit power budget, the maximum \cgls{tp} increases. Instead, the \cgls{gee} saturates at some point and the transmit power stays constant in the \cgls{gee} optimal resource allocation. Increasing the transmit power beyond the \cgls{gee} saturation point, as is done in the \cgls{tp} optimal allocation, decreases the \cgls{gee}. For a maximum transmit power of \SI{23}{\dBm}, which corresponds to the typical \cgls{ue} power budget \cite{Dahlman2018,Joshi2017},
the \gls{gee} optimal allocation achieves \SI{22.4}{\percent} or \SI{0.84}{\mega\bit\per\second} less \cgls{tp} than possible. Instead, the \cgls{htee} resource allocation is within \SI{95}{\percent} of the maximum achievable \cgls{tp} and achieves a \SI{97}{\percent} higher \cgls{gee} than the maximum \cgls{tp} allocation at \SI{23}{\dBm}. This corresponds to a gain of \SI{1.8}{\mega\bit\per\joule} at the cost of \SI{0.19}{\mega\bit\per\second}.

\tikzsetnextfilename{TP}
\tikzpicturedependsonfile{throughput.dat}

\tikzsetnextfilename{GEE}
\tikzpicturedependsonfile{gee.dat}

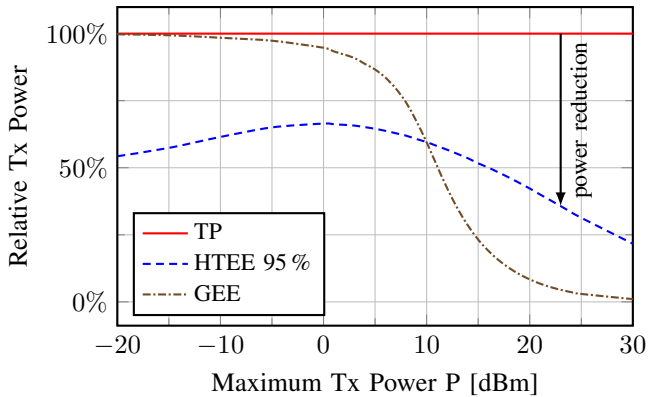
\begin{figure}
\centering
\begin{tikzpicture}
	\begin{axis} [
			thick,
			ylabel near ticks,
			xlabel near ticks,
			grid=both,
			minor x tick num = 1,
			minor y tick num = 1,
			no markers,
			legend style={font=\small},
			legend pos=south west,
			legend cell align=left,
			xlabel={Maximum Tx Power P [dBm]},
			ylabel={Relative Tx Power},
			yticklabel = {$\pgfmathprintnumber{\tick}\%$},
			xmin = -20,
			xmax = 30,
			width=\axisdefaultwidth,
			height=.8*\axisdefaultheight,
		]

		\pgfplotstableread[col sep=comma]{powVSrate.dat}\tbl

		\addplot+[smooth] table[y=TIN_Prel] {\tbl};
		\addlegendentry{TP};

		\addplot+[smooth] table[y=HTEE95_Prel] {\tbl};
		\addlegendentry{HTEE \SI{95}{\percent}};

		\addplot+[smooth] table[y=GEE_Prel] {\tbl};
		\addlegendentry{GEE};

		\draw [latex-] (axis cs:23,35.69160909895825) -- node [sloped,below,font=\small] {power reduction} (axis cs:23,100);
	\end{axis}
\end{tikzpicture}
\vspace{-2ex}
\caption{Total power consumption relative to the throughput optimal strategy.}
\label{fig:prel}
\end{figure}

However, the \cgls{gee} is not the optimal metric to evaluate transmit power savings. Consider a second operating point at \SI{-10}{\dBm}, the median transmit power of 4G \cglspl{ue} in urban scenarios \cite{Joshi2017}. The \cgls{tp} and \cgls{gee} optimal strategies both achieve almost the same \cgls{tp} and \cgls{gee}. \Cref{fig:prel} displays the power consumption relative to the \cgls{tp} optimal resource allocation. It can be observed that the \cgls{gee} strategy consumes almost as much transmit power as the \cgls{tp} strategy, and, thus, is unable to exploit the ``rate reduction budget'' of the system designer.
Instead, the \cgls{htee} strategy uses almost \SI{40}{\percent} less transmit power at a \cgls{tp} cost of \SI{50}{\kilo\bit\per\second}, which is less than the data rate of classical digital telephone line modem. Nevertheless, its \cgls{gee} is worse than that of the other strategies, despite the tremendous transmit power reduction.

\begin{figure}
\centering
\begin{tikzpicture}
	\begin{axis} [
			thick,
			ylabel near ticks,
			xlabel near ticks,
			grid=both,
			minor x tick num = 1,
			minor y tick num = 1,
			no markers,
			legend style={font=\small},
			legend pos=south west,
			legend cell align=left,
			xlabel={Throughput [Mbit/s]},
			ylabel={Relative Tx Power},
			yticklabel = {$\pgfmathprintnumber{\tick}\%$},
			width=\axisdefaultwidth,
			height=.8*\axisdefaultheight,
		]

		\pgfplotstableread[col sep=comma]{powVSrate.dat}\tbl

		\addplot+[smooth] table[x=TIN, y=TIN_Prel] {\tbl};
		\addlegendentry{TP};

		\addplot+[smooth] table[x=HTEE95, y=HTEE95_Prel] {\tbl};
		\addlegendentry{HTEE \SI{95}{\percent}};

		\addplot+[smooth] table[x=GEE, y=GEE_Prel] {\tbl};
		\addlegendentry{GEE};

		\draw [-latex] (axis cs:3.7494322092643406,100.0) -- node [sloped,below,font=\small] {power--rate trade-off} (axis cs:3.562832369897267,35.69160909895825);
		\draw [-latex] (axis cs:3.7494322092643406,100.0) -- (axis cs:2.9080868711476136,4.489341114237462);
	\end{axis}
\end{tikzpicture}
\vspace{-2ex}
\caption{Relative power consumption as a function of the achievable throughput.}
\label{fig:prelrate}
\vspace{-2ex}
\end{figure}
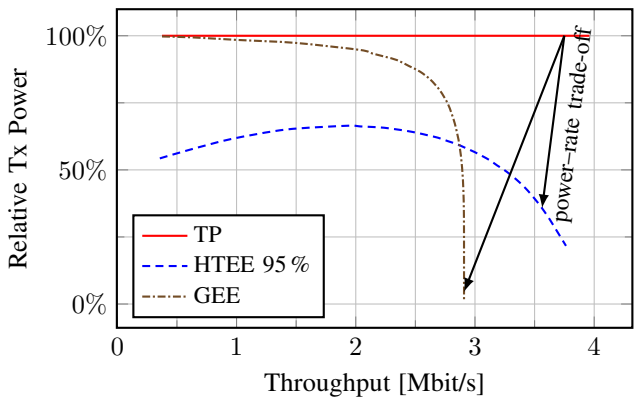

Returning to our previous scenario with \SI{23}{\dBm} maximum transmit power, it can be seen from \cref{fig:prel} that
the \cgls{htee} strategy consumes only \SI{35.7}{\percent} of the transmit power necessary to achieve the maximum \cgls{tp}. Of course the \cgls{gee} strategy saves even more transmit power but at a much higher cost to the throughput.
This trade-off is illustrated in \cref{fig:prelrate} where the relative transmit power is plotted over the achievable data rate.
It can be observed that a major advantage of the \cgls{htee} strategy over the \cgls{gee} optimal power allocation is that the \cgls{tp} does not saturate and any data rate is achievable given a sufficient transmit power budget. Thus, it results in an energy-efficient resource allocation while still ensuring high \cgls{tp}.

\tikzsetnextfilename{prel}
\tikzpicturedependsonfile{powVSrate.dat}

\tikzsetnextfilename{prelrate}
\tikzpicturedependsonfile{powVSrate.dat}

Finally, 
to support the statement at the end of \cref{sec:mmp} that \cref{opt:pmin} is hard to solve with a traditional \cgls{bb} method, we have also employed the \cgls{mmp} framework to solve \cref{opt:pmin}. Out of 1000 problem instances that ran on an Intel Xeon E5-2680 v3 CPU with a memory usage limit of \SI{21}{\giga\byte}, 483 problem instances ran out of memory and 517 problem instances did not complete within 24 hours, i.e., not a single problem instance of \cref{opt:pmin} could be solved by a traditional \cgls{bb} algorithm with reasonable usage of computational resources. In contrast, the same problem instances could be solved with \cref{alg:sit} using a maximum of \SI{50}{\mega\byte} memory and not taking longer than \SI{752}{\milli\second} to complete. The median computation time among all problem instances was  \SI{1.75}{\milli\second}.

\section{Conclusions}
We have introduced the novel concept of hierarchical resource allocation and applied it to minimize energy consumption while still ensuring high spectrum utilization. The numerical results show a transmit power reduction of \SI{65}{\percent} in a multi-cell communication scenario at the cost of a \SI{5}{\percent} drop in \cgls{tp}. Instead, state-of-the-art \cgls{gee} maximization results in a \cgls{tp} reduction of almost \SI{25}{\percent}. Moreover, this strategy also saves energy in scenarios where \cgls{gee} optimization fails to provide a gain over \cgls{tp} maximization.
The developed algorithms solve the involved optimization problems with global optimality and, therefore, rigorously demonstrate the gains of hierarchical resource allocation and high-throughput energy efficiency maximization over state-of-the-art approaches.

\balance
\bibliography{IEEEabrv,IEEEtrancfg,conf}

\begin{thebibliography}{10}
\providecommand{\url}[1]{#1}
\csname url@samestyle\endcsname
\providecommand{\newblock}{\relax}
\providecommand{\bibinfo}[2]{#2}
\providecommand{\BIBentrySTDinterwordspacing}{\spaceskip=0pt\relax}
\providecommand{\BIBentryALTinterwordstretchfactor}{4}
\providecommand{\BIBentryALTinterwordspacing}{\spaceskip=\fontdimen2\font plus
\BIBentryALTinterwordstretchfactor\fontdimen3\font minus
  \fontdimen4\font\relax}
\providecommand{\BIBforeignlanguage}[2]{{%
\expandafter\ifx\csname l@#1\endcsname\relax
\typeout{** WARNING: IEEEtran.bst: No hyphenation pattern has been}%
\typeout{** loaded for the language `#1'. Using the pattern for}%
\typeout{** the default language instead.}%
\else
\language=\csname l@#1\endcsname
\fi
#2}}
\providecommand{\BIBdecl}{\relax}
\BIBdecl

\bibitem{Han2008}
Z.~Han and K.~J.~R. Liu, \emph{Resource Allocation for Wireless Networks:
  Basics, Techniques, and Applications}.\hskip 1em plus 0.5em minus 0.4em\relax
  Cambridge Univ. Press, 2008.

\bibitem{Zhang2017}
H.~Zhang, N.~Liu, X.~Chu, K.~Long, A.~Aghvami, and V.~C.~M. Leung, ``Network
  slicing based {5G} and future mobile networks: Mobility, resource management,
  and challenges,'' \emph{{IEEE} Commun. Mag.}, vol.~55, no.~8, pp. 138--145,
  Aug. 2017.

\bibitem{Popovski2018}
P.~Popovski, K.~F. Trillingsgaard, O.~Simeone, and G.~Durisi, ``{5G} wireless
  network slicing for {eMBB}, {URLLC}, and {mMTC}: A communication-theoretic
  view,'' \emph{{IEEE} Access}, vol.~6, pp. 55\,765--55\,779, Sep. 2018.

\bibitem{Weeraddana2012}
P.~C. Weeraddana, M.~Codreanu, M.~Latva-aho, A.~Ephremides, and C.~Fischione,
  \emph{Weighted Sum-Rate Maximization in Wireless Networks: A Review}, ser.
  FnT Netw.\hskip 1em plus 0.5em minus 0.4em\relax Now, 2012, vol.~6, no. 1-2.

\bibitem{Grandhi1993}
S.~A. Grandhi, R.~Vijayan, D.~J. Goodman, and J.~Zander, ``Centralized power
  control in cellular radio systems,'' \emph{{IEEE} Trans. Veh. Technol.},
  vol.~42, no.~4, pp. 466--468, Nov. 1993.

\bibitem{Isheden2012}
C.~Isheden, Z.~Chong, E.~Jorswieck, and G.~Fettweis, ``Framework for link-level
  energy efficiency optimization with informed transmitter,'' \emph{{IEEE}
  Trans. Wireless Commun.}, pp. 1--12, 2012.

\bibitem{Zappone2015}
A.~Zappone and E.~Jorswieck, \emph{Energy Efficiency in Wireless Networks via
  Fractional Programming Theory}, ser. FNT in Communications and Information
  Theory.\hskip 1em plus 0.5em minus 0.4em\relax Now Publishers, 2015, vol.~11,
  no. 3-4.

\bibitem{Zadeh1963}
L.~A. Zadeh, ``Optimality and non-scalar-valued performance criteria,''
  \emph{{IEEE} Trans. Autom. Control}, vol.~8, no.~1, pp. 59--60, Jan. 1963.

\bibitem{Zhang2010a}
R.~Zhang and S.~Cui, ``Cooperative interference management with {MISO}
  beamforming,'' \emph{{IEEE} Trans. Signal Process.}, vol.~58, no.~10, pp.
  5450--5458, Oct. 2010.

\bibitem{Aydin2017}
O.~Aydin, E.~A. Jorswieck, D.~Aziz, and A.~Zappone, ``Energy-spectral
  efficiency tradeoffs in {5G} multi-operator networks with heterogeneous
  constraints,'' \emph{{IEEE} Trans. Wireless Commun.}, vol.~16, no.~9, pp.
  5869--5881, Sep. 2017.

\bibitem{Miettinen1999}
K.~Miettinen, \emph{Nonlinear Multiobjective Optimization}.\hskip 1em plus
  0.5em minus 0.4em\relax Springer, 1999.

\bibitem{Dahlman2018}
E.~Dahlman, S.~Parkvall, and J.~Sk\"old, \emph{{5G} {NR}: The Next Generation
  Wireless Access Technology}, 1st~ed.\hskip 1em plus 0.5em minus 0.4em\relax
  Academic Press, 2018.

\bibitem{satwp}
I.~Leyva-Mayorga \emph{et~al.}, ``{LEO} small-satellite constellations for {5G}
  and beyond-{5G} communications,'' \emph{{IEEE} Access}, vol.~8, Oct. 2020.

\bibitem{Zappone2016}
A.~Zappone, L.~Sanguinetti, G.~Bacci, E.~Jorswieck, and M.~Debbah,
  ``Energy-efficient power control: A look at {5G} wireless technologies,''
  \emph{{IEEE} Trans. Signal Process.}, vol.~64, no.~7, pp. 1668--1683, Apr.
  2016.

\bibitem{Bestle1997}
D.~Bestle and P.~Eberhard, ``Dynamic system design via multicriteria
  optimization,'' in \emph{Multiple Criteria Decision Making: Proceedings of
  the Twelth International Conference}, ser. Lecture Notes in Economics and
  Mathematical Systems, G.~Fandel and T.~Gal, Eds.\hskip 1em plus 0.5em minus
  0.4em\relax Springer, 1997.

\bibitem{Luo2008}
Z.-Q. Luo and S.~Zhang, ``Dynamic spectrum management: Complexity and
  duality,'' \emph{{IEEE} J. Sel. Areas Commun.}, vol.~2, no.~1, Feb. 2008.

\bibitem{mmp}
B.~Matthiesen, C.~Hellings, E.~A. Jorswieck, and W.~Utschick, ``Mixed monotonic
  programming for fast global optimization,'' \emph{{IEEE} Trans. Signal
  Process.}, vol.~68, pp. 2529--2544, Mar. 2020.

\bibitem{Tuy2000}
H.~Tuy, ``Monotonic optimization: Problems and solution approaches,''
  \emph{SIAM J. Optimization}, vol.~11, no.~2, pp. 464--494, Feb. 2000.

\bibitem{Tuy2009}
------, ``{$\mathcal{D(C)}$}-optimization and robust global optimization,''
  \emph{J. Global Optim.}, vol.~47, no.~3, pp. 485--501, Oct. 2009.

\bibitem{Tuy2016}
------, \emph{Convex Analysis and Global Optimization}.\hskip 1em plus 0.5em
  minus 0.4em\relax Springer, 2016.

\bibitem{tsp2018}
B.~Matthiesen and E.~A. Jorswieck, ``Efficient global optimal resource
  allocation in non-orthogonal interference networks,'' \emph{{IEEE} Trans.
  Signal Process.}, vol.~67, no.~21, pp. 5612--5627, Nov. 2019.

\bibitem{HataCOST231}
3GPP, ``Digital cellular telecommunications systems (phase 2+); radio network
  planning aspects,'' Tech. Rep. TR 43.030 V9.0.0 R9, Feb. 2010.

\bibitem{Rappaport2002}
T.~S. Rappaport, \emph{Wirless Communications}, 2nd~ed.\hskip 1em plus 0.5em
  minus 0.4em\relax Prentice-Hall, 2002.

\bibitem{Joshi2017}
P.~Joshi, D.~Colombi, B.~Thors, L.-E. Larsson, and C.~T\"ornevik, ``Output
  power levels of {4G} user equipment and implications on realistic {RF} {EMF}
  exposure assessments,'' \emph{{IEEE} Access}, vol.~5, pp. 4545--4550, Mar.
  2017.

\end{thebibliography}
\end{document}